\documentclass[onecolumn,12pt]{IEEEtran}

\usepackage{amssymb}
\usepackage{graphicx}
\usepackage{amsmath}
\usepackage[english]{babel}
\usepackage{mathrsfs}

\newtheorem{thm}{Theorem}

\newtheorem{definition}{Definition}
\newtheorem{proposition}[thm]{Proposition}
\newtheorem{corollary}[thm]{Corollary}

\newtheorem{example}{Example}

\begin{document}
\title {Guessing Revisited: A Large Deviations Approach}
\author{Manjesh Kumar Hanawal and Rajesh Sundaresan,~\IEEEmembership{Senior Member,~IEEE}
\thanks{This work was supported by the Defence Research and Development Organisation, Ministry of Defence, Government of India, under the DRDO-IISc Programme on Advanced Research in Mathematical Engineering, and by the University Grants Commission under Grant Part (2B) UGC-CAS-(Ph.IV).}
\thanks{The material in this paper was presented in part at the IEEE International Symposium on Information Theory
(ISIT 2007) held in Nice, France, June 2007, in part at the IISc Centenary Conference on Managing Complexity in a Distributed World, (MCDES 2008) held in Bangalore, India, May 2008, and in part at the National Conference on Communications (NCC 2009), Guwahati, India, Jan. 2009.
}
}

\maketitle

\begin{abstract}
The problem of guessing a random string is revisited. A close relation between guessing and compression is first established. Then it is shown that if the sequence of distributions of the information spectrum satisfies the large deviation property with a certain rate function, then the limiting guessing exponent exists and is a scalar multiple of the Legendre-Fenchel dual of the rate function. Other sufficient conditions related to certain continuity properties of the information spectrum are briefly discussed. This approach highlights the importance of the information spectrum in determining the limiting guessing exponent. All known prior results are then re-derived as example applications of our unifying approach.
\end{abstract}

\begin{IEEEkeywords}
guessing, length function, source coding, information spectrum, large deviations.
\end{IEEEkeywords}

\section{Introduction}
\label{sec:introduction}
Let $X^n= (X_1,\cdots,X_n)$ denote $n$ letters of a process where each letter is drawn from a finite set $\mathbb{X}$ with joint probability mass function (pmf) $(P_n(x^n):x^n \in \mathbb{X}^n)$. Let $x^n$ be a realization and suppose that we wish to guess this realization by asking questions of the form ``Is $X^n = x^n$?'', stepping through the elements of $\mathbb{X}^n$ until the answer is ``Yes''. We wish to do this using the minimum expected number of guesses. There are several applications that motivate this problem. Consider cipher systems employed in digital television or DVDs to block unauthorized access to special features. The ciphers used are amenable to such exhaustive guessing attacks and it is of interest to quantify the effort needed by an attacker (Merhav \& Arikan \cite{199909TIT_MerAri}).

Massey \cite{1994_ProcISIT_Mas} observed that the expected number of guesses is minimized by guessing in the decreasing order of $P_n$-probabilities. Define the {\em guessing function} $$G_n^*: \mathbb{X}^n \rightarrow \{ 1, 2, \cdots, |\mathbb{X}|^n \}$$ to be one such optimal guessing order\footnote{If there are several sequences with the same probability of occurrence, they may be guessed in any order without affecting the expected number of guesses.}. $G_n^*(x^n) = g$ implies that $x^n$ is the $g$th guess. Arikan \cite{Arikan} considered the growth of $\mathbb{E} \left[ G_n^*(X^n)^{\rho} \right]$ as a function of $n$ for an independent and identically distributed (iid) source with marginal pmf $P_1$ and $\rho > 0$. He showed that the growth is exponential in $n$; the limiting exponent
\begin{equation}
  \label{eqn:limitingExponent}
  E(\rho) := \lim_{n \rightarrow \infty}\frac{1}{n}\ln \mathbb{E}[G_n^*(X^n)^{\rho}]
\end{equation}
exists and equals $\rho H_{\alpha}(P_1)$ with $\alpha = 1/(1+\rho)$, where $H_{\alpha}(P_n)$ is the R\'{e}nyi entropy of order $\alpha$ for the pmf $P_n$, given by
\begin{equation}
  \label{eqn:RE}
  \frac{1}{1-\alpha} \ln \left( \sum_{x^n \in \mathbb{X}^n} P_n(x^n)^{\alpha} \right), ~\alpha \neq 1.
\end{equation}
Malone \& Sullivan \cite{200403TIT_MalSul} showed that the limiting exponent $E(\rho)$ of an irreducible Markov chain exists and equals the logarithm of the {\em Perron-Frobenius eigenvalue} of a matrix formed by raising each element of the transition probability matrix to the power $\alpha$. From their proof, one obtains the more general result that the limiting exponent exists for any source if the R\'{e}nyi entropy {\em rate} of order $\alpha$,
\begin{equation}
  \label{eqn:RER}
  \lim_{n \rightarrow \infty} n^{-1} H_{\alpha}(P_n),
\end{equation}
exists for $\alpha = 1/(1+\rho)$. Pfister \& Sullivan \cite{200411TIT_PfiSul} showed the existence of (\ref{eqn:limitingExponent}) for a class of stationary probability measures, beyond Markov measures, that are supported on proper subshifts of $\mathbb{X}^{\mathbb{N}}$ \cite{200411TIT_PfiSul}. A particular example is that of shifts generated by finite-state machines. For such a class, they showed that the guessing exponent has a variational characterization (see (\ref{eqn:Erho-iid}) later). For unifilar sources Sundaresan \cite{200805MCDES_Sun} obtained a simplification of this variational characterization using a direct approach and the method of types.

Merhav \& Arikan remark that their proof in \cite{Arikan-Merhav} for the limiting guessing exponent is equally applicable to finding the limiting exponent of the moment generating function of compression lengths. Moreover, the two exponents are the same. The latter is a problem studied by Campbell \cite{Campbell-1}.

Our contribution is to give a large deviations perspective to these results, shed further light on the aforementioned connection between compression and guessing, and unify all prior results on existence of limiting guessing exponents. Specifically, we show that
if the sequence of distributions of the {\em information spectrum} $(1/n) \ln (1 / P_n(X^n))$ (see Han \cite{2003ISMIT_Han}) satisfies the {\em large deviation property}, then the limiting exponent exists. This is useful because several existing large deviations results can be readily applied. We then show that all but one previously considered cases in the literature\footnote{These are cases without side information and key-rate constraints. The one exception is an example of Arikan \& Merhav \cite[Sec. VI-B]{Arikan-Merhav} for which one can show the existence of R\'{e}nyi entropy rate rather directly via a subadditivity argument. See our technical report \cite{200812TRPME_HanSun}.} satisfy this sufficient condition. See Examples \ref{example:iid}-\ref{example:Mixed-Source} in section \ref{sec:examples}.

The large deviation theoretic ideas are already present in the works of Pfister \& Sullivan \cite{200411TIT_PfiSul} and the method of types approach of Arikan \& Merhav \cite{Arikan-Merhav}. Our work however brings out the essential ingredient (the sufficient conditions on the information spectrum), and enables us to see the previously obtained specific results under one light.

The quest for a general sufficient condition under which the information spectrum satisfies a large deviation property is a natural line of inquiry, and one of independent interest, in view of the Shannon-McMillan-Breiman theorem which asserts that the information spectrum of a stationary and ergodic source converges to the Shannon entropy almost surely and in $L_q$, for all $q \geq 1$; see for example \cite{2007xxCTCQIT_Par}. In particular, the large deviation property implies exponentially fast convergence to entropy. In the several specific examples we consider, the information spectrum does satisfy the large deviation property. One sufficient condition for the weaker property of exponentially fast convergence to entropy is the so-called {\em blowing up property}. (See Marton \& Shields \cite[Th. 2]{1994xxIJM_MarShi}, or the survey article by Shields \cite{199810TIT_Shi}). One family of sources, that includes most of the sources we consider in this paper and goes beyond, is that of {\em finitary encodings} of memoryless processes, also called finitary processes. These are known to have the blowing-up property, and therefore exponentially fast convergence to entropy (see Marton \& Shields \cite[Th. 3]{1994xxIJM_MarShi}). It is an interesting open question to see if finitary processes, or what other sources with the blowing up property, satisfy the large deviation property.

The rest of the paper is organized as follows. Section II studies the tight relationship between guessing and compression. Section III states the relevant large deviations results and the main sufficiency results. Section IV re-derives prior results by showing that in each case the information spectrum satisfies the LDP. Section V contains proofs and section VI contains some concluding remarks.


\section{Guessing and Compression}

\label{sec:GuessCodeEquivalence}

In this section we relate the problem of guessing to one of source compression. An interesting conclusion is that
robust source compression strategies lead to robust guessing strategies.

For ease of exposition, let us assume that the message space is simply $\mathbb{X}$. The
extension to strings of length $n$ is straightforward and will be returned to shortly.
A guessing function
\[
  G : \mathbb{X} \rightarrow \left\{1, 2, \cdots, |\mathbb{X}| \right\}
\]
is a bijection that denotes the order in which the elements of
$\mathbb{X}$ are guessed. If $G(x) = g$, then the $g$th guess is
$x$. Let $\mathbb{N}$ denote the set of natural numbers. A length function
\[
  L : \mathbb{X} \rightarrow \mathbb{N}
\]
is one that satisfies Kraft's inequality
\begin{equation}
  \label{eqn:KraftInequality}
  \sum_{x \in \mathbb{X}} \exp_2\{-L(x)\} \leq 1,
\end{equation}
where we have used the notation $\exp_2\{-L(x)\} = 2^{-L(x)}$.
To each guessing function $G$, we associate a PMF $Q_G$ on
$\mathbb{X}$ and a length function $L_G$ as follows.

\begin{definition}
Given a guessing function $G$, we say $Q_G$ defined by
\begin{equation}
  \label{eqn:Q_G}
  Q_G(x) = c^{-1} \cdot G(x)^{-1}, ~\forall x \in \mathbb{X},
\end{equation}
is the PMF on $\mathbb{X}$ associated with $G$. The quantity $c$ in
(\ref{eqn:Q_G}) is the normalization constant. We say $L_G$ defined
by
\begin{equation}
  \label{eqn:L_G}
  L_G(x) = \left\lceil - \log_2 Q_G(x) \right\rceil, ~\forall x \in
  \mathbb{X},
\end{equation}
is the length function associated with $G$.
\end{definition}

Observe that
\begin{equation}
  \label{eqn:c}
  c = \sum_{a \in \mathbb{X}} G(a)^{-1} = \sum_{i=1}^{|\mathbb{X}|} \frac{1}{i} \leq 1 + \ln
  |\mathbb{X}|,
\end{equation}
and therefore the PMF in (\ref{eqn:Q_G}) is well-defined. We record
the intimate relationship between these associated quantities in the
following result. (This is also available in the proof of
\cite[Th. 1, p.382]{199203TIT_WeiZivLem}).

\begin{proposition}
\label{prop:guessingBounds} Given a guessing function $G$, the
associated quantities satisfy
\begin{eqnarray}
  \label{eqn:guessing-PMFBounds}
  c^{-1} \cdot Q_G(x)^{-1} =
  G(x) \leq Q_G(x)^{-1}, \\
  \label{eqn:guessing-lengthBounds}
  L_G(x) - 1 - \log_2 c \leq \log_2 G(x) \leq L_G(x).
\end{eqnarray}
\end{proposition}
\begin{IEEEproof}
The first equality in (\ref{eqn:guessing-PMFBounds}) follows from
the definition in (\ref{eqn:Q_G}), and the second inequality from
the fact that $c \geq 1$.

The upper bound in (\ref{eqn:guessing-lengthBounds}) follows from
the upper bound in (\ref{eqn:guessing-PMFBounds}) and from
(\ref{eqn:L_G}). The lower bound in
(\ref{eqn:guessing-lengthBounds}) follows from
\begin{eqnarray*}
  \log_2 G(x) & = & \log_2 \left( c^{-1} \cdot Q_G(x)^{-1} \right) \\
  & = & - \log_2 Q_G(x) - \log_2 c \\
  & \geq & \left( \lceil - \log_2 Q_G(x) \rceil - 1 \right) - \log_2 c
  \\
  & = & L_G(x) - 1 - \log_2 c.
\end{eqnarray*}
\end{IEEEproof}

We now associate a guessing function $G_L$ to each length function
$L$.

\begin{definition} \label{defn:G_L}
Given a length function $L$, we define the associated guessing
function $G_L$ to be the one that guesses in the increasing order of
$L$-lengths. Messages with the same $L$-length are ordered using an
arbitrary fixed rule, say the lexicographical order on $\mathbb{X}$.
We also define the associated PMF $Q_L$ on $\mathbb{X}$ to be
\begin{equation}
  \label{eqn:Q_L}
  Q_L(x) = \frac{\exp_2\{-L(x)\}}{\sum_{a \in \mathbb{X}}
  \exp_2\{-L(a)\}}.
\end{equation}
\end{definition}

\begin{proposition}
\label{prop:lengthBounds} For a length function $L$, the associated
PMF and the guessing function satisfy the following:
\begin{enumerate}
\item $G_L$ guesses messages in the decreasing order of $Q_L$-probabilities;
\item
\begin{equation}
  \label{eqn:G_LBounds}
  \log_2 G_L(x) \leq \log_2 Q_L(x)^{-1} \leq L(x).
\end{equation}
\end{enumerate}
\end{proposition}

\begin{IEEEproof}
The first statement is clear from the definition of $G_L$ and from
(\ref{eqn:Q_L}).

Letting $1 \{ E \}$ denote the indicator function of an event $E$,
we have as a consequence of statement 1) that
\begin{eqnarray}
  G_L(x) \nonumber
  & \leq & \sum_{a \in \mathbb{X}} 1 \left\{ Q_L(a) \geq Q_L(x) \right\} \nonumber \\
  & \leq & \sum_{a \in \mathbb{X}} \frac{Q_L(a)}{Q_L(x)} \nonumber \\
  \label{eqn:guessingUpperBound}
  & = & Q_L(x)^{-1},
\end{eqnarray}
which proves the left inequality in (\ref{eqn:G_LBounds}). This
inequality was known to Wyner \cite{197203IC_Wyn}.

The last inequality in (\ref{eqn:G_LBounds}) follows from
(\ref{eqn:Q_L}) and Kraft's inequality (\ref{eqn:KraftInequality})
as follows:
\[
  Q_L(x)^{-1} = \exp_2\{L(x)\} \cdot \sum_{a \in
  \mathbb{X}} \exp_2\{-L(a)\} \leq \exp_2\{L(x)\}.
\]
\end{IEEEproof}

Let $\{L(x) \geq B \}$ denote the set $\{ x \in \mathbb{X} \mid L(x)
\geq B \}$. We then have the following easy to verify corollary to
Propositions \ref{prop:guessingBounds} and \ref{prop:lengthBounds}.

\begin{corollary}
\label{cor:inclusions} For a given $G$, its associated length
function $L_G$, and any $B \geq 1$, we have
\begin{eqnarray}
  \lefteqn{ \left\{ L_G(x)  \geq B + 1 + \log_2 c \right\} } \nonumber \\
  & & \subseteq \left\{ G(x) \geq \exp_2\{B\} \right\} \nonumber \\
  \label{eqn:containments}
  & & \subseteq \left\{ L_G(x) \geq B \right\}.
\end{eqnarray}

Analogously, for a given $L$, its associated guessing function
$G_L$, and any $B \geq 1$, we have
\begin{equation}
  \label{eqn:GsubsetL}
  \{ G_L(x) \geq \exp_2\{B\}\} \subseteq \{ L(x) \geq B\}.
\end{equation}
\end{corollary}
The inequalities between the associates in
(\ref{eqn:guessing-lengthBounds}) and (\ref{eqn:G_LBounds}) indicate
the direct relationship between guessing moments and Campbell's
coding problem \cite{Campbell-1}, and that the R\'{e}nyi entropies
are the optimal growth exponents for guessing moments, as highlighted
in the following Proposition.

\begin{proposition}
Let $L$ be any length function on $\mathbb{X}$, $G_{L}$ the guessing
function associated with $L$, $P$ a PMF on $\mathbb{X}$, $\rho \in
(0, \infty)$, $L^*$ the length function that minimizes $\mathbb{E}
\left[ \exp_2\{\rho L^*(X)\} \right]$, where the expectation is with
respect to $P$, $G^*$ the guessing function that proceeds in the
decreasing order of $P$-probabilities and therefore the one that
minimizes $\mathbb{E} \left[ G^*(X)^{\rho} \right]$, and $c$ as in
(\ref{eqn:c}). Then
\begin{equation}
  \label{eqn:ratioUpperBound}
  \frac{\mathbb{E} \left[ G_{L}(X)^{\rho} \right] }{\mathbb{E} \left[ G^*(X)^{\rho} \right]}
  \leq \frac{\mathbb{E} \left[ \exp_2\{\rho L(X)\} \right] }{\mathbb{E} \left[ \exp_2\{\rho L^*(X)\}
  \right]} \cdot \exp_2\{\rho ( 1 + \log_2 c )\}.
\end{equation}
Analogously, let $G$ be any guessing function, and $L_G$ its
associated length function. Then
\begin{equation}
  \label{eqn:ratioLowerBound}
  \frac{\mathbb{E} \left[ G(X)^{\rho} \right] }{\mathbb{E} \left[ G^*(X)^{\rho} \right]}
  \geq \frac{\mathbb{E} \left[ \exp_2\{\rho L_G(X)\} \right] }{\mathbb{E} \left[ \exp_2\{\rho L^*(X)\}
  \right]} \cdot \exp_2\{- \rho ( 1 + \log_2 c )\}.
\end{equation}
Also,
\begin{equation}
  \label{eqn:G*L*Relation}
  \left| \frac{1}{\rho} \log_2 \mathbb{E} \left[ G^*(X)^{\rho} \right]
  - \frac{1}{\rho} \log_2 \mathbb{E} \left[ \exp_2\{\rho L^*(X)\}
  \right] \right| \leq 1 + \log_2 c.
\end{equation}
\end{proposition}

\begin{IEEEproof}
Observe that
\begin{eqnarray}
  \lefteqn{ \mathbb{E} \left[ \exp_2\{\rho L(X)\} \right] }
  \nonumber \\
  \label{eqn:2a}
  & \geq & \mathbb{E} \left[ G_{L}(X)^{\rho}
  \right] \\
  & \geq & \mathbb{E} \left[ G^*(X)^{\rho} \right] \nonumber \\
  \label{eqn:2b}
  & \geq & \mathbb{E} \left[ \exp_2\{\rho L_{G^*}(X)\} \right] \exp_2\{-\rho ( 1 + \log_2 c )\} \\
  \label{eqn:2c}
  & \geq & \mathbb{E} \left[ \exp_2\{ \rho L^*(X)\} \right] \exp_2\{ -\rho ( 1 + \log_2 c )\},
\end{eqnarray}
where (\ref{eqn:2a}) follows from (\ref{eqn:G_LBounds}), and
(\ref{eqn:2b}) from the left inequality in
(\ref{eqn:guessing-lengthBounds}). The result in
(\ref{eqn:ratioUpperBound}) immediately follows. A similar argument
shows (\ref{eqn:ratioLowerBound}). Finally, (\ref{eqn:G*L*Relation})
follows from the inequalities leading to (\ref{eqn:2c}) by setting
$L = L^*$.
\end{IEEEproof}

Thus if we have a length function whose performance is close to
optimal, then its associated guessing function is close to guessing
optimal. The converse is true as well. Moreover, the optimal
guessing exponent is within $1+ \log_2 c$ of the optimal coding
exponent for the length function.

\subsection{Strings of length $n$}
Let us now consider strings of length $n$. Let $\mathbb{X}^n$ denote
the set of messages and consider $n \rightarrow \infty$. Let $\mathcal{M}(\mathbb{X}^n)$ denote the set of pmfs on $\mathbb{X}^n$. By a source, we mean a sequence of pmfs $(P_n : n \in \mathbb{N})$, where $P_n \in \mathcal{M}(\mathbb{X}^n)$.
We replace the normalization constant $c$ in (7) by $c_n$ and observe that
\[
c_n \leq 1 + n \ln |\mathbb{X}|.
\]
If we normalize both sides of equation (\ref{eqn:G*L*Relation}) by $n$,
the difference between two quantities as a function of $n$ decays as $O((\log_2 n) / n)$, and vanishes as $n$ tends to infinity. The following theorem follows immediately, with a change of base to natural logarithms.

\begin{thm}
\label{thm:GuessCodeLimit}
Given $\rho >0$, the limit
\[
\lim_{n \rightarrow \infty }n^{-1}\ln \mathbb{E}[G_n^*(X^n)^{\rho}]
\] exists if and only if the limit
\[
\lim_{n \rightarrow \infty}\inf_{L_n}n^{-1}\ln \mathbb{E} [\exp_2 \{\rho L_n(X^n )\}]
\]
exists. Furthermore, the two limits are equal.
\end{thm}

\vspace*{.15in}

It is therefore sufficient to restrict our attention to the Campbell's coding problem \cite{Campbell-1} and study if the limit
\begin{equation}
  \label{eqn:Campbell}
  \lim_{n \rightarrow \infty}\inf_{L_n}\frac{1}{n}\ln \mathbb{E} [\exp \{(\rho \ln 2) L_n(X^n )\}]
\end{equation}
exists, where the infimum is taken over all length functions $L_n:\mathbb{X}^n \rightarrow \mathbb{N}$ and exponentiation is with respect to the base of the natural logarithm.

\subsection{Universality}
Before we proceed to studying the limit, we make a further remark on the connection between {\em universal}
strategies for guessing and universal strategies for compression.

Let $\mathbb{T}$ denote a class of sources. For each source in the
class, let $P_n$ be its restriction to strings of length $n$ and let
$L_n^*$ denote an optimal length function that attains the minimum
value $\mathbb{E} \left[ \exp\{(\rho \ln 2) L_n^*(X^n)\}\right]$ among all
length functions, the expectation being with respect to $P_n$. On
the other hand, let $L_n$ be a sequence of length functions for the
class of sources that does not depend on the actual source within
the class. Suppose further that the length sequence $L_n$ is
asymptotically optimal, i.e.,
\begin{eqnarray*}
  \lefteqn{ \lim_{n \rightarrow \infty} \frac{1}{n \rho} \ln \mathbb{E} \left[ \exp\{(\rho \ln 2) L_n(X^n)\} \right] } \\
  & = & \lim_{n \rightarrow \infty} \frac{1}{n \rho} \ln \mathbb{E} \left[ \exp\{(\rho \ln 2) L_n^*(X^n)\}
  \right],
\end{eqnarray*}
for every source belonging to the class. $L_n$ is thus ``univeral''
for (i.e., asymptotically optimal for all sources in) the class. An
application of (\ref{eqn:ratioUpperBound}) with $c_n$ in place of $c$
followed by the observation $(1 + \log_2 c_n)/n \rightarrow 0$ shows that the sequence of guessing
strategies $G_{L_n}$ is asymptotically optimal for the class, i.e.,
\begin{eqnarray}
  \lefteqn{ \lim_{n \rightarrow \infty} \frac{1}{n \rho} \ln \mathbb{E} \left[ G_{L_n}(X^n)^{\rho} \right] } \nonumber \\
  \label{eqn:growthExponent}
  & = & \lim_{n \rightarrow \infty} \frac{1}{n \rho} \ln \mathbb{E} \left[ G^*(X^n)^{\rho} \right] \nonumber.
\end{eqnarray}

Arikan and Merhav \cite{Arikan-Merhav} provide a universal guessing
strategy for the class of discrete memoryless sources (DMS). For the
class of unifilar sources with a known number of states, the minimum
description length encoding is asymptotically optimal for Campbell's
coding length problem (see Merhav \cite{199105TIT_Mer}). It follows
as a consequence of the above argument that guessing in the
increasing order of description lengths is asymptotically optimal.
The left side of (\ref{eqn:ratioUpperBound}) is the extra factor in the
expected number of guesses (relative to the optimal value) due to
lack of knowledge of the specific source in class. Sundaresan \cite{200701TIT_Sun}
characterized this loss as a function of the
uncertainty class.


\section{Large Deviation Results}
\label{sec:LDResults}

We begin with some words on notation. Recall that $\mathcal{M}(\mathbb{X}^n)$ denotes the set of pmfs on $\mathbb{X}^n$. The Shannon entropy for a $P_n \in \mathcal{M}(\mathbb{X}^n)$ is
\[
  H(P_n)=-\sum_{x^n \in \mathbb{X}^n} P_n(x^n) \ln P_n(x^n)
\]
and the R\'{e}nyi entropy of order $\alpha \neq 1$ is (\ref{eqn:RE}). The Kullback-Leibler divergence or relative entropy between two pmfs $Q_n$ and $P_n$ is
\[
  D(Q_n \parallel P_n) = \left\{
                       \begin{array}{ll}
                         \hspace*{-.1in} \displaystyle \sum_{x^n \in \mathbb{X}^n } \hspace*{-.05in}Q_n(x^n) \ln  \frac{Q_n(x^n)}{P_n(x^n)}, & \mbox{if } Q_n \ll P_n, \\
                         & \\
                         \infty, & \mbox{otherwise,}
                       \end{array}
  \right.
\]
where $Q_n \ll P_n$ means $Q_n$ is absolutely continuous with respect to $P_n$. Recall that a source is a sequence of pmfs $(P_n : n \in \mathbb{N})$ where $P_n \in \mathcal{M}(\mathbb{X}^n)$. It is usually obtained via $n$-length marginals of some probability measure in $\mathcal{M}(\mathbb{X}^{\mathbb{N}})$. Also recall the definitions of limiting guessing exponent in (\ref{eqn:limitingExponent}) and R\'{e}nyi entropy rate in (\ref{eqn:RER}) when the limits exist. $G_n^*$ is an optimal guessing function for a pmf $P_n \in \mathcal{M}(\mathbb{X}^n)$. From the results in Section \ref{sec:GuessCodeEquivalence} on the equivalence between guessing and compression, it is sufficient to focus on the Campbell coding problem.

Our first contribution is a proof of the following implicit result of Malone \& Sullivan \cite{200403TIT_MalSul}. The proof is given in Section \ref{subsec:prop1}.
\vspace*{.1in}
\begin{proposition}
\label{prop:renyi-guessing-equivalence}
Let $\rho > 0$. For a source $(P_n : n \in \mathbb{N})$, $E(\rho)$ exists if and only if the R\'{e}nyi entropy rate (\ref{eqn:RER}) exists. Furthermore, $E(\rho) / \rho$ equals the R\'{e}nyi entropy rate.
\end{proposition}
\vspace*{.1in}
The question now boils down to the existence of the limit in the definition of R\'{e}nyi entropy rate. The theory of large deviations immediately yields a sufficient condition. We begin with a definition.

\vspace*{.1in}
\begin{definition}[Large deviation property]\cite[Def. II.3.1]{1985ELDSM_Ellis}
\label{defn:ldp}
A sequence $(\nu_n : n \in \mathbb{N})$ of probability measures on $\mathbb{R}$ satisfies the {\em large deviation property (LDP)} with rate function $I:\mathbb{R} \rightarrow [0, \infty]$ if the following conditions hold:
\begin{itemize}
  \item $I$ is lower semicontinuous on $\mathbb{R}$;
  \item $I$ has compact level sets;
  \item $\limsup_{n \rightarrow \infty} n^{-1} \ln \nu_n\{K\} \leq - \inf_{t \in K} I(t)$ for each closed subset $K$ of $\mathbb{R}$;
  \item $\liminf_{n \rightarrow \infty} n^{-1} \ln \nu_n\{G\} \geq - \inf_{t \in G} I(t)$ for each open set $G$ of $\mathbb{R}$.
\end{itemize}
\end{definition}
\vspace*{.1in}

Several commonly encountered sources satisfy the LDP with known and well-studied rate functions. We describe some of these in the examples treated subsequently.

Let $\nu_n$ denote the distribution of the information spectrum given by the real-valued random variable $-n^{-1} \ln P_n(X^n)$.
The following proposition gives a sufficient condition for the existence of the limiting R\'{e}nyi entropy rate (and therefore the limiting guessing exponent).

\vspace*{.1in}
\begin{proposition}
  \label{prop:ldpSufficientCondition}
  Let the sequence of distributions $(\nu_n : n \in \mathbb{N})$ of the information spectrum satisfy the LDP with rate function $I$. Then the limiting R\'{e}nyi entropy rate of order $1/(1+\rho)$ exists for all $\rho > 0$ and equals
  \[
    \beta^{-1} \sup_{t \in \mathbb{R}} ~ \{\beta t - I(t)\},
  \]
  where $\beta= \rho/(1+\rho)$. Consequently, the limiting guessing exponent exists and equals
  \[
    (1+\rho) \sup_{t \in \mathbb{R}} ~ \{\beta t - I(t)\}.
  \]
\end{proposition}
\vspace*{.1in}

The function $I^*(\beta) := \sup_{t \in \mathbb{R}} ~ \{\beta t - I(t)\}$ is the Legendre-Fenchel dual of the rate function $I$. Proposition \ref{prop:ldpSufficientCondition} says that, under the sufficient condition, the limiting guessing exponent equals $(1+\rho) I^*(\rho / (1+\rho))$, and is thus directly related to the large deviations rate function for information spectrum. This is however different from Merhav \& Arikan's \cite[Th. 2]{Arikan-Merhav} for memoryless sources which states that the limiting guessing exponent is the Legendre-Fenchel dual of the source coding {\em error exponent} function. We refer the reader to Merhav and Arikan \cite[Sec. IV]{Arikan-Merhav} for further interesting connections between source coding error exponent, guessing exponent, and two other exponents related to lossy source coding.

Let us briefly discuss another approach to verify the existence of R\'{e}nyi entropy rate (see Proposition \ref{prop:renyi-guessing-equivalence}). With $\alpha = 1/(1+\rho)$, we can rewrite $1 - \alpha$ times the R\'{e}nyi entropy rate in (\ref{eqn:RER}) as
\begin{eqnarray}
  \lefteqn{ (1 - \alpha) \lim_{n \rightarrow \infty} n^{-1} H_{\alpha}(P_n) } \nonumber \\
  \label{eqn:normalizedpartition}
  & = & \hspace*{-.1in} \lim_{n \rightarrow \infty} n^{-1} \ln \sum_{x^n \in \mathbb{X}^n} \exp \left\{ - n \alpha F_n(x^n) \right \} U_n(x^n),
\end{eqnarray}
where
$$ F_n(x^n) := \left( - n^{-1}\ln P_{n}(x^n) - (\ln |\mathbb{X}|) / \alpha \right), $$
and $U$ is the iid process on $X^{\mathbb{N}}$ with uniform marginal on $\mathbb{X}$. One can then view $\alpha \in (0,1)$ as the inverse temperature (when $\rho > 0$) of a statistical mechanical system, $F_n(x^n)$ as the energy of the configuration $x^n$, and the right side of (\ref{eqn:normalizedpartition}) as a scaled version of (i.e., $\alpha$ times) the specific Gibbs free energy of the corresponding statistical mechanical system, if the limit exists. This view point is particularly useful because the iid process $U$ satisfies a sample path large deviation property. If the information spectrum sequence satisfies the continuity conditions in Varadhan \cite[Th. 3.4]{1966xxCPAM_Var}, then the limiting specific Gibbs free energy exists, and so does the R\'{e}nyi entropy rate. Our technical report \cite{200812TRPME_HanSun} treats an example via this more general approach.

\subsection{Additional results from Large Deviations Theory}
In order to study the examples in Section \ref{sec:examples}, we state some additional results on LDP of transformed variables.
  (See \cite[Sec. 4.2]{1998LDTA_DemZei}), \cite[Th. 6.12 and 6.14]{200610LN-LDASM_Ell}).

\begin{proposition}[Contraction Principle]
 \label{prop:ContractionPrinciple}
Let $(\xi_n : n \in \mathbb{N})$ denote a sequence of $\mathcal{X}$-valued random variables where $\mathcal{X}$ is a complete separable metric space (Polish space). Let $\nu_n$ denote the distribution of $\xi_n$ for $n \in \mathbb{N}$, and let the sequence of distributions $(\nu_n : n \in \mathbb{N})$ on $\mathcal{X}$ satisfy the LDP with rate function $I: \mathcal{X} \rightarrow [0,\infty]$. Let $\phi:\mathcal{X} \rightarrow \mathbb{R}$ be a continuous function. The sequence of distributions of $(\phi(\xi_n): n \in \mathbb{N})$ on $\mathbb{R}$ also satisfies the LDP with rate function $J: \mathbb{R} \rightarrow [0, \infty]$ given by

 \[
J(y)=\inf \{I(x) : x \in \mathbb{R}, \phi(x)=y  \}.
\]
\end{proposition}

\vspace*{.15in}

\begin{proposition}[Exponential Approximation]
\label{prop:Exponential Approximation}
Suppose that the sequence of distributions of $(\xi_n : n \in \mathbb{N})$ satisfies the LDP with rate function $I$ on $\mathbb{R}$. Assume also that the sequence of random variables $(\zeta_n : n \in \mathbb{N})$ is superexponentially close  to  $(\xi_n : n \in \mathbb{N})$ in the following sense: for each $\delta > 0$
\begin{equation}
\label{eqn:superexponential}
\limsup_{n \rightarrow \infty}\frac{1}{n}\ln \Pr\{|\xi_n - \zeta_n|>\delta\}= - \infty.
\end{equation}
Then the sequence of distributions of $(\zeta_n : n \in \mathbb{N})$ also satisfies the LDP on $\mathbb{R}$ with the same rate function $I$. The condition in (\ref{eqn:superexponential}) is satisfied if
\begin{equation}
\label{eqn:limitsupremumapprox}
 \lim_{n \rightarrow \infty}\sup_{\omega \in \Omega} \left|\xi_n(\omega)-\zeta_n(\omega)\right|=0,
\end{equation}
where $\Omega$ is the underlying sample space.
\end{proposition}


\section{Examples}
\label{sec:examples}

We are now ready to apply Proposition 7 and related techniques to various examples. In first five examples that follow, our goal is to show that the sufficient condition for the existence of the limiting guessing exponent holds, i.e., that the sequence of distributions of the information spectrum satisfies the LDP.

\subsection{LDP for information spectrum}

\begin{example}[An iid source]
\label{example:iid}
This example was first studied by Arikan \cite{Arikan}.
Recall that an iid source is one for which $P_n(x^n) = \prod_{i=1}^n P_1(x_i)$, where $P_1$ is the marginal of $X_1$. It is then clear that the  information spectrum can be written as a sample mean of iid random variables
\[
  - n^{-1} \ln P_n(X^n) = - n^{-1} \sum_{i=1}^n \ln P_1(X_i).
\]
It is well-known that the sequence $(\nu_n : n \in \mathbb{N})$ of distributions of this sample mean satisfies the LDP with rate function given by the Legendre-Fenchel dual of the cumulant of the random variable $- \ln P_1(X_1)$ (see for example \cite[Th. II.4.1]{1985ELDSM_Ellis} or \cite[eqn. (1.9.66-67)]{2003ISMIT_Han}):
\begin{eqnarray*}
  \ln \mathbb{E} \bigg[ \exp \Big\{ \beta (- \ln P_1(X_1)) \Big\} \bigg] & = & \ln \left( \sum_{x \in \mathbb{X}} P_1(x)^{\alpha} \right) \\
  & = &(1 - \alpha) H_{\alpha}(P_1).
\end{eqnarray*}
The Legendre-Fenchel dual of the rate function is therefore the cumulant itself (\cite[Th. VI.4.1.e]{1985ELDSM_Ellis}). An application of Proposition \ref{prop:ldpSufficientCondition} yields that $(1+\rho)$ times this cumulant, given by $\rho H_{\alpha}(P_1)$, is the guessing exponent. We thus recover Arikan's result \cite{Arikan}.

The rate function $I$ can also be obtained using the {\em contraction principle}  (Proposition \ref{prop:ContractionPrinciple}) as follows. This method will provide a recipe to obtain the limiting guessing exponent in subsequent examples. Consider a mapping that takes $x^n$ to its empirical pmf in $\mathcal{M}(\mathbb{X})$. Empirical pmf is then a random variable. The distribution of $X^n$ induces a pmf on $\mathcal{M}(\mathbb{X})$. It is well-known that the sequence of distributions of these empirical pmfs, indexed by $n$, satisfies the {\em level-2} LDP\footnote{Level-1 refers to sequence of distributions (indexed by $n$) of sample means, level-2 refers to sample histograms, and level-3 to sample paths.} with rate function $I^{(2)}_{P_1}(\cdot) = D(\cdot \parallel P_1)$. See for example \cite[Th II.4.3]{1985ELDSM_Ellis}. Observe that the mapping from the empirical pmf to the information spectrum random variable is continuous. We can therefore use the contraction principle to get a formula for $I$ in terms of $I^{(2)}_{P_1}(\cdot)$ as follows \cite[Th II.5.1]{1985ELDSM_Ellis}. For any $t$ in $\mathbb{R}$, let
\[
  \theta(t) := \Big\{ Q \in \mathcal{M}(\mathbb{X}) : \sum_{x \in \mathbb{X}} Q(x)\ln \frac{1}{P_1(x)} = t \Big\},
\]
i.e.,
\[
  \theta(t) = \Big\{ Q \in \mathcal{M}(\mathbb{X}) : H(Q) + D(Q \parallel P_1) = t \Big\}.
\]
Then
\[
  I(t)= \inf\{I^{(2)}_{P_1}(Q):Q \in \theta(t)\}.
\]
Using this, we can write
\begin{eqnarray*}
I^*(\beta) & = & \displaystyle \sup_{t \in \mathbb{R}} \Big\{\beta t - \inf_{Q \in \theta (t)}D(Q \parallel P_1)\Big\}  \\
  & =  & \sup_{t \in \mathbb{R}} \sup_{Q \in \theta (t)} \Big\{\beta t - D(Q \parallel P_1)\Big\} \\
  & =  & \hspace*{-0.05in} \sup_{Q \in \mathcal{M}(\mathbb{X})} \Big\{ \beta (H(Q)+ D(Q \parallel P_1)) - D(Q \parallel P_1)\Big\} \\
  & =  & (1+\rho)^{-1} \sup_{Q \in \mathcal{M}(\mathbb{X})} \Big\{\rho H(Q) - D(Q \parallel P_1)\Big\},
\end{eqnarray*}
thus yielding
\begin{equation}
  \label{eqn:Erho-iid}
  E(\rho) = \sup_{Q \in \mathcal{M}(\mathbb{X})} \Big\{\rho H(Q) - D(Q \parallel P_1)\Big\}.
\end{equation}
This formula extends to more general sources, as is seen in the next few examples.
\end{example}

\begin{example}[Markov source]
\label{example:Markov}
This example was studied by Malone \& Sullivan \cite{200403TIT_MalSul}. Consider an irreducible Markov chain taking values on $\mathbb{X}$ with transition probability matrix $\pi$. Our goal is to verify that the sufficient condition holds and to calculate $E(\rho)$ defined by (\ref{eqn:limitingExponent}) for this source.

Let $\mathcal{M}_s(\mathbb{X}^2)$ denote the set of {\em stationary} pmfs defined by
\begin{eqnarray*}
  \lefteqn{ \mathcal{M}_s \left(\mathbb{X}^2\right) = \Big\{ Q \in \mathcal{M} \left(\mathbb{X}^2\right) : } \\
  && ~~~~~~~~~~~~\sum_{x_1 \in \mathbb{X}} Q(x_1, x) = \sum_{x_2 \in \mathbb{X}} Q(x,x_2) \forall x \in \mathbb{X} \Big\}.
\end{eqnarray*}
Denote the common marginal by $q$ and let
\[
  \eta(\cdot \mid x_1) := \left\{
      \begin{array}{cl}
        Q(x_1, \cdot) / q(x_1), &  \mbox{ if } q(x_1) \neq 0, \\
        1/|\mathbb{X}|, & \mbox{otherwize}.
      \end{array}
  \right.
\]
We may then denote $Q = q \times \eta$, where $q$ is the distribution of $X_1$ and $\eta$ the conditional distribution of $X_2$ given $X_1$. It is once again well known that the empirical pmf random variable satisfies the level-2 LDP with rate function $I^{(2)}_{\pi}(Q)$, given by \cite{198505TIT_Nat}
\begin{eqnarray*}
  I^{(2)}_{\pi}(Q) & = & D(\eta \parallel \pi \mid q) \\
  & := & \sum_{x_1 \in \mathbb{X}} q(x_1) D(\eta(\cdot \mid x_1) \parallel \pi(\cdot \mid x_1)).
\end{eqnarray*}
As in Example \ref{example:iid}, the contraction principle then yields that the sequence of distributions of information spectrum satisfies the LDP with rate function $I$ given by
\[
  I(t)= \displaystyle\inf\{I^{(2)}_{\pi}(Q): Q \in \theta (t)\}.
\]
where for $t$ in $\mathbb{R}$,  $\theta(t) \subset \mathcal{M}_s(\mathbb{X}^2)$ is defined by
\[
  \theta(t)=\left\{ Q \in \mathcal{M}_s(\mathbb{X}^2) : \sum_{x_1,x_2} Q(x_1,x_2) \ln \frac{1}{\pi(x_2|x_1)}=t \right \}.
\]
By Proposition \ref{prop:renyi-guessing-equivalence}, the limiting guessing exponent exists. Perron-Frobenius theory (Seneta \cite[Ch. 1]{1973xxNNM_Sen}, see also \cite[pp.60-61]{2000LD_FDH}) yields the cumulant directly as $\ln \lambda(\beta)$, where $\lambda(\beta)$ is unique largest eigenvalue (Perron-Frobenius eigenvalue) of a matrix formed by raising each element of $\pi$ to the power $\alpha$. (Recall that $\alpha = 1/(1+\rho)$ and $\beta = \rho / (1+\rho)$). Thus $E(\rho) = (1+\rho) \ln \lambda(\beta)$, and we recover the result of Malone \& Sullivan \cite{200403TIT_MalSul}. It is useful to note that the steps that led to (\ref{eqn:Erho-iid}) hold in the Markov case (with appropriate changes to entropy and divergence terms) and we may write
\begin{equation}
  \label{eqn:Erho-Markov}
  E(\rho) = \sup_{Q \in \mathcal{M}_s(\mathbb{X}^2)} \Big\{\rho H(\eta \mid q) - D(\eta \parallel \pi \mid q)\Big\},
\end{equation}
where $H(\eta \mid q)$ is the conditional entropy of $X_2$ given $X_1$ under the joint distribution $Q$, i.e.,
\[
  H(\eta \mid q) := - \sum_{x \in \mathbb{X}} q(x) H(\eta(\cdot \mid x)).
\]
\end{example}

\begin{example}[Unifilar source]
\label{example:unifilar}
This example was studied by Sundaresan in \cite{200805MCDES_Sun}. A unifilar source is a generalization of the Markov source in Example \ref{example:Markov}. Let $\mathbb{X}$ denote the alphabet set as before. In addition, let $\mathbb{S}$ denote a set of finite states. Fix an initial state $s_0$ and let the joint probability of observing $(x^n, s^n)$ be
\[
  P_n(x^n, s^n) = \prod_{i=1}^n \pi(x_i, s_i \mid s_{i-1})
\]
where $\pi(x_i, s_i \mid s_{i-1})$ is the joint probability of $(x_i, s_i)$ given the previous state $s_{i-1}$. The dependence of $P_n$ on $s_0$ is understood. Furthermore, assume that $\pi(x_i, s_i \mid s_{i-1})$ is such that $s_i = \phi(s_{i-1},x_i)$, where $\phi$ is a deterministic function that is one-to-one for each fixed $s_{i-1}$. Such a source is called a unifilar source.

$P_{S,X}(s_{i-1},x_i)$ and $\phi$ completely specify the process: the initial state $S_0$ is random with distribution that of marginal of $S$ in $P_{S,X}$, the rest being specified by $P_{X|S}(x_i \mid s_{i-1})$ and $\phi$. Example \ref{example:Markov} is a unifilar source with $\mathbb{S} = \mathbb{X}$, $\phi(s_{i-1},x_i) = x_i$, and $P_{S,X} = q \times \pi$ where $q$ is the stationary distribution of the Markov chain.

Let $\mathcal{M}_s(\mathbb{S} \times \mathbb{X})$ denote the set of joint measures on the indicated space so that the resulting process $(S_n : n \geq 0)$ is a stationary and irreducible Markov chain. Let a $Q \in \mathcal{M}_s (\mathbb{S} \times \mathbb{X})$ be written as $Q = q \times \eta$. For any $t$ in $\mathbb{R}$, let
\[
  \theta(t) := \left\{ Q \in \mathcal{M}_s (\mathbb{S} \times \mathbb{X}) : \sum_{(s,x)} Q(s,x) \ln \frac{1}{\pi(x \mid s)} = t \right\}.
\]
Then the sequence of distributions of information spectrum $- n^{-1} \ln P_n(X^n)$ satisfies the LDP (\cite[eqn. (1.9.30)]{2003ISMIT_Han}) with rate function given (once again via contraction principle) by
\[
  I(t) = \inf \{ D(\eta \parallel \pi \mid q) : Q \in \theta(t)\}.
\]
The limiting exponent therefore exists. Following the same procedure that led to (\ref{eqn:Erho-iid}) in the iid case and (\ref{eqn:Erho-Markov}) for a Markov source, we get
\begin{equation}
  \label{eqn:Erho-unifilar}
  E(\rho) = \sup_{Q \in \mathcal{M}_s(\mathbb{S} \times \mathbb{X})} \Big\{\rho H(\eta \mid q) - D(\eta \parallel \pi \mid q)\Big\},
\end{equation}
where $H(\eta \mid q)$ and $D(\eta \parallel \pi \mid q)$ are analogously defined, and the result of Sundaresan \cite{200805MCDES_Sun} is recovered.
\end{example}

\begin{example}[A class of stationary sources]
\label{example:stationary}
Pfister \& Sullivan \cite{200411TIT_PfiSul} considered a class of stationary sources with distribution $P \in \mathcal{M}\left(\mathbb{X}^{\mathbb{N}}\right)$ that satisfies two hypotheses H1 and H2 of \cite[Sec. II-B]{200411TIT_PfiSul}, which we will now describe.

Let $\mathcal{M}^P(\mathbb{X}^{\mathbb{N}})$ denote the set of sources that satisfy $Q_n \ll P_n$ for all $n \in \mathbb{N}$, where $P_n$ and $Q_n$ are restrictions of $P$ and $Q$ to $n$ letters. Note that it may be possible that a $Q \in \mathcal{M}^P(\mathbb{X}^{\mathbb{N}})$ is not absolutely continuous with respect to $P$. Also, let $\mathcal{M}^P_s(\mathbb{X}^{\mathbb{N}}) \subset \mathcal{M}^P(\mathbb{X}^{\mathbb{N}})$ denote the subset of stationary sources with respect to the shift operator $\tau: \mathbb{X}^\mathbb{N} \rightarrow  \mathbb{X}^\mathbb{N}$ defined by
\[
(\tau(x))_i = x_{i+1}, \forall i \in \mathbb{N}.
\]
Hypothesis H1 of Pfister \& Sullivan \cite{200411TIT_PfiSul} assumes that for any neighborhood of a stationary source $Q \in \mathcal{M}^P_s(\mathbb{X}^{\mathbb{N}})$ and any $\varepsilon > 0$, there exists an ergodic $Q' \in \mathcal{M}^P_s(\mathbb{X}^{\mathbb{N}})$ in that neighborhood such that $\overline{H}(Q') \geq \overline{H}(Q) - \varepsilon$, where $\overline{H}(Q)$ is the Shannon entropy rate of source $Q$. Their hypothesis H2 is given by (\ref{eqn:expodifference}) below.

Under these hypotheses, Pfister \& Sullivan \cite{200411TIT_PfiSul} proved that $E(\rho)$ exists, and provided a variational characterization analogous to (\ref{eqn:Erho-unifilar}), i.e.,
\begin{equation}
  \label{eqn:Erho-stationary}
  E(\rho) = \sup_{Q \in {\mathcal{M}}_s^{P}(\mathbb{X}^{\mathbb{N}})} \Big\{ \rho \overline{H}(Q) - \overline{D}(Q \parallel P) \Big\},
\end{equation}
where
\[
  \overline{D}( Q \parallel P) = \lim_{n \rightarrow \infty} n^{-1} \sum_{x^n} Q_n(x^n) \ln \frac{Q_n(x^n)}{P_n(x^n)}.
\]

En route to this result, Pfister \& Sullivan \cite{200411TIT_PfiSul} showed that the sequence of distributions of the {\em empirical process} satisfies the {\em level-3} LDP for sample paths. We first state this precisely, and then use this as the starting point to show the sufficient condition that the information spectrum satisfies the LDP.

For an $x \in \mathbb{X}^\mathbb{N}$ given by $x = (x_1, x_2, \cdots)$, we define $x^n = (x_1, \cdots, x_n)$ as the first $n$ components of $x$ in the usual way. Consider a stationary source $P$ whose letters are $X = (X_1, X_2, \cdots)$. Define the empirical process of measures
\[
T_n(X, \cdot) = n^{-1} \sum_{i=0}^{n-1} \delta_{\tau^i(X)}(\cdot).
\]
This is a measure on $\mathbb{X}^\mathbb{N}$ that puts mass $1/n$ on the following strings: $x, \tau(x), \tau^2(x), \cdots, \tau^{n-1}(x)$. Pfister \& Sullivan showed that the distributions of the $\mathcal{M}(\mathbb{X}^{\mathbb{N}})$-valued process $T_n(X, \cdot)$ satisfies the level-3 LDP with rate function $I_P^{(3)}(\cdot) = \overline{D}(\cdot \parallel P)$ under hypotheses H1 and H2 of their paper (\cite[Prop. 2.2-2.3]{200411TIT_PfiSul}). Furthermore,
\begin{equation}
\label{eqn:restrictiontostationary}
  \overline{D}(Q \parallel P) = + \infty, \quad Q \notin \mathcal{M}^P_s(\mathbb{X}^{\mathbb{N}}),
\end{equation}
so that we may restrict $\overline{D}(\cdot \parallel P)$ to $\mathcal{M}^P_s(\mathbb{X}^{\mathbb{N}})$. We next use this to show that the information spectrum satisfies the LDP.

Hypothesis H2 of Pfister \& Sullivan assumes the existence of a continuous mapping $e_P : \mathbb{X}^\mathbb{N} \rightarrow \mathbb{R}$ satisfying
\begin{equation}
\label{eqn:expodifference}
\lim_{n \rightarrow \infty} \sup_{x \in \Sigma_n^P} \left |n^{-1}\ln P_n(x^n)+\int_{\mathbb{X}^\mathbb{N}}e_P~  dT_n(x,\cdot)~\right |=0,
\end{equation}
where $\Sigma_n^P = \{ x \in \mathbb{X}^{\mathbb{N}} : P_n(x^n) > 0\}$.

By the compactness of  $\mathbb{X}^\mathbb{N}$, $e_P$ is uniformly continuous. Under the weak topology on the complete separable metric space $\mathcal{M}( \mathbb{X}^\mathbb{N})$, the mapping
\[
\phi : \mathcal{M}(\mathbb{X}^\mathbb{N}) \rightarrow \mathbb{R}
\]
defined by $Q \mapsto \int_{\mathbb{X}^\mathbb{N}} e_P ~ dQ$ is a continuous mapping. Hence by the contraction principle, by setting $\mathcal{X} = \mathcal{M}(\mathbb{X}^{\mathbb{N}})$ we get that the sequence of distributions of $(\phi(T_n(X, \cdot) : n \in \mathbb{N})$ satisfies the LDP with rate function $I$ given by
\[
I(t)=\inf\left \{\overline{D}( Q \parallel P) : Q \in \mathcal{M}_s^{P}(\mathbb{X}^{\mathbb{N}}), \phi(Q)=t \right\},
\]
where the restriction of the infimum to $\mathcal{M}^P_s(\mathbb{X}^{\mathbb{N}})$ follows from (\ref{eqn:restrictiontostationary}). Furthermore, given hypothesis H2 and (\ref{eqn:expodifference}), an application of the exponential approximation principle (Proposition \ref{prop:Exponential Approximation}) indicates that the sequence of distributions of the information spectrum too satisfies the LDP with the same rate function $I$, and we have verified that the sufficient condition holds.

What remains is to calculate this rate function. For this, we return to Pfister \& Sullivan's work and use
$\overline{D}(Q \parallel P)=\phi(Q)-\overline{H}(Q)$ \cite[Prop. 2.1]{200411TIT_PfiSul} to write
\[
 I(t)=\inf_{Q \in \mathcal{M}_s^{P}}\left \{\overline{D}( Q \parallel P) :\overline{H}(Q) + \overline{D}(Q \parallel P)=t \right\}.
\]

Finally, the Legendre-Fenchel dual of the rate function is computed as in the steps leading to (\ref{eqn:Erho-iid})-(\ref{eqn:Erho-unifilar}), yielding (\ref{eqn:Erho-stationary}).

\end{example}

\begin{example}[Mixed source]
\label{example:Mixed-Source}
Consider a mixture of two iid sources with letters from $\mathbb{X}$. We may write
\[
  P_n(x^n)= \lambda \prod_{i=1}^{n}R(x_i) + (1-\lambda)\prod_{i=1}^{n}S(x_i)
\]
where $\lambda \in (0,1)$ with $R,S \in \mathcal{M}(\mathbb{X})$ the two marginal pmfs that define the iid components of the mixture.
It is easy to see that the guessing exponent is the maximum of the guessing exponents for the two component sources. We next verify this using Proposition \ref{prop:ldpSufficientCondition}.

The sequence of distributions of the information spectrum satisfies the LDP with rate function given as follows (see Han \cite[eqn. (1.9.41)]{2003ISMIT_Han}). Define
\begin{eqnarray*}
  \theta_1 & = & \Big\{Q \in \mathcal{M}(\mathbb{X}) : D(Q \parallel S)-D(Q \parallel R) \geq 0  \Big \}, \\
  \theta_2 & = & \Big \{Q \in \mathcal{M}(\mathbb{X}) : D(Q \parallel S)-D(Q \parallel R) \leq 0 \Big \},
\end{eqnarray*}
and for $t \in \mathbb{R}$
\begin{eqnarray*}
A_t & = & \theta_1 \cap \Big \{Q \in \mathcal{M}(\mathcal{\mathbb{X}}) : H(Q)+D(Q \parallel R) =t \Big \} \\
B_t & = & \theta_2 \cap \Big \{Q \in \mathcal{M}(\mathbb{X}) : H(Q)+D(Q \parallel S) =t \Big \}.
\end{eqnarray*}
The rate function (via the contraction principle) is given by
\[I(t)=\min \left \{\displaystyle \inf_{Q \in A_t} D(Q \parallel R),  \displaystyle \inf_{Q \in B_t} D(Q \parallel S)\right \}.\]
From Proposition \ref{prop:ldpSufficientCondition} we conclude that the limiting guessing exponent exists. $I^*(\beta)$ is then
\begin{eqnarray*}
\lefteqn{\sup_{t \in \mathbb{R}} \bigg \{ \beta t -  \min \Big \{ \inf_{Q \in A_t} D(Q \parallel R), \inf_{Q \in B_t} D(Q \parallel S)\Big\} \bigg \} }\\
&=&\max \bigg \{ \lefteqn{\sup_{t \in \mathbb{R}} \sup_{Q \in A_t}\Big \{\beta t -  D(Q \parallel R)\Big \},}\\
&&\hspace{0.4in}\sup_{t \in \mathbb{R}} \sup_{Q \in B_t}\Big \{\beta t - D(Q \parallel S)\Big \} \bigg\} \\
&=&\max \bigg \{ \sup_{Q \in \theta_1 }\Big \{\beta H(Q) -(1-\beta)  D(Q \parallel R)\Big \}, \\
& & \hspace{0.4in} \sup_{Q \in \theta_2}\Big \{\beta H(Q) -(1-\beta) D(Q \parallel S)\Big \} \bigg\} \\
& = & (1+ \rho)^{-1}  \max \bigg \{  \sup_{Q} \lefteqn{ \Big \{\rho H(Q)-D(Q \parallel R)\Big \} ,} \\
   && \hspace{1.0in} \sup_{Q}\Big\{\rho H(Q)-D(Q \parallel S)\Big \}\bigg \}  \\
   &=& (1+ \rho)^{-1}\max \Big \{\rho H_{\alpha}(R),\rho H_{\alpha}(S) \Big \},
\end{eqnarray*}
yielding \[E(\rho)=\max \Big\{\rho H_{\alpha}(R),\rho H_{\alpha}(S) \Big\}.\]
\end{example}


\section{Proofs}
\label{sec:proofs}
We now prove Propositions \ref{prop:renyi-guessing-equivalence} and \ref{prop:ldpSufficientCondition}.

\subsection{Proof of Proposition \ref{prop:renyi-guessing-equivalence}}
\label{subsec:prop1}
From Theorem \ref{thm:GuessCodeLimit} it is sufficient to show that the limit in (\ref{eqn:Campbell}) for Campbell's coding problem exists if and only if the R\'{e}nyi entropy rate exists, with the former $\rho$ times the latter.

Fix $n$. In the rest of the proof, we use the notation $\mathbb{E}_{P_n}[\cdot]$ for expectation with respect to distribution $P_n$. The length function can be thought of as a bounded (continuous) function from $\mathbb{X}^n$ to $\mathbb{R}$ and therefore our interest is in the logarithm of its moment generating function of $\rho$, the cumulant. The cumulant associated with a bounded continuous function (here $L_n$) has a variational characterization \cite[Prop. 1.4.2]{1999AWCATLDP_DupEll} as the following Legendre-Fenchel dual of the Kullback-Leibler divergence, i.e.,
\begin{eqnarray}
  \lefteqn{ \ln \mathbb{E}_{P_n}\Big [\exp \{ (\rho \ln 2) L_n(X^n)\}\Big ] } \nonumber \\
     & = & \sup_{Q_n \in \mathcal{M}(\mathbb{X}^n)}\Big \{ (\rho \ln 2) \mathbb{E}_{Q_n}[L_n(X^n)]-D(Q_n \parallel P_n)\Big\}. \nonumber \\
     \label{eqn:legendreTransform}
\end{eqnarray}
Taking infimum on both sides over all length functions, we arrive at the following chain of inequalities:
\begin{eqnarray}
  \label{eqn:unnormalizedCampbellEqn}
  \lefteqn{ \displaystyle \inf_{L_n}\ln \mathbb{E}_{P_n}\Big [\exp \{(\rho \ln 2) L_n(X^n)\}\Big ] } \\
   & = & \displaystyle \inf_{L_n}\displaystyle \sup_{Q_n \in \mathcal{M}(\mathbb{X}^n)}\Big \{\mathbb{E}_{Q_n}[(\rho \ln 2) L_n(X^n)]-D(Q_n \parallel P_n)\Big \} \nonumber \\
     & = & \lefteqn {\hspace*{-.1in}\sup_{Q_n \in \mathcal{M}(\mathbb{X}^n)}   \inf_{L_n}  \Big \{\mathbb{E}_{Q_n}[(\rho \ln 2) L_n(X^n)]-D(Q_n \parallel P_n)\Big \} } \nonumber\\
 \label{eqn:sup-inf}
 && \hspace{2in}+ \Theta(1)\\
   \label{eqn:entropy}
   & = &  \hspace*{-.1in}\sup_{Q_n \in \mathcal{M}(\mathbb{X}^n)} \Big \{ \rho H_n(Q_n) - D(Q_n \parallel P_n)\Big \} + \Theta(1) \\
   \label{eqn:variational}
   & = & \rho H_{\frac{1}{1+ \rho}}(P_n) + \Theta(1).
\end{eqnarray}
Equation (\ref{eqn:sup-inf}) follows because (i) the mapping $$(L_n, Q_n) \mapsto \mathbb{E}_{Q_n}[(\rho \ln 2) L_n(X^n)]-D(Q_n \parallel P_n)$$ is a concave function of $Q_n$; (ii) for fixed $Q_n$ and for any two length functions  $L_n^{(1)}$ and $L_n^{(2)}$, for any $\lambda \in [0,1]$, the function
$$L_n=\left \lceil \lambda L_n^{(1)} + (1-\lambda)L_n^{(2)} \right \rceil$$
is also a length function and
\[\mathbb{E}_{Q_n}[L_n]= \lambda \mathbb{E}_{Q_n}[L_n^{(1)}]+(1-\lambda) \mathbb{E}_{Q_n}[L_n^{(2)}]+ \Theta(1);\]
(iii) $\mathcal{M}(\mathbb{X}^n)$ is compact and convex, and therefore the infimum and supremum may be interchanged upon an application of a version of Ky Fan's minimax result \cite{198204_MATHACADHUNG}. This yields a compression problem, the infimum over $L_n$ of expected lengths with respect to a distribution $Q_n$. The answer is the well-known Shannon entropy $H(Q_n)$ to within $\ln 2$ nats, and (\ref{eqn:entropy}) follows. Lastly, (\ref{eqn:variational}) is a well-known identity which may also be obtained directly by writing the supremum term in (\ref{eqn:entropy}) as
\begin{eqnarray*}
(1+\rho) \sup_{Q_n \in \mathcal{M}(\mathbb{X}^n)} \Big\{ \mathbb{E}_{Q_n} \left[ - \left(\frac{\rho}{1+\rho} \right) \ln P_n(X^n) \right] \\
 - ~ D(Q_n \parallel P_n) \Big\}
\end{eqnarray*}
and then applying (\ref{eqn:legendreTransform}) with $-(\rho/(1+\rho) \ln P_n(X^n))$ in place of $(\rho \ln 2) L_n(X^n)$ to get the scaled R\'{e}nyi entropy.

Normalize both (\ref{eqn:unnormalizedCampbellEqn}) and (\ref{eqn:variational}) by $n$ and let $n \rightarrow \infty$ to deduce that (\ref{eqn:Campbell}) exists if and only if the limiting normalized R\'{e}nyi entropy rate exists. This concludes the proof.

\subsection{Proof of Proposition \ref{prop:ldpSufficientCondition}}
\label{subsec:prop2}
This is a straightforward application of Varadhan's theorem \cite{1966xxCPAM_Var} on asymptotics of integrals. Recall that $\nu_n$ is the distribution of the information spectrum $n^{-1} \ln P_n(X^n)$. Define $F(t) = \beta t$. Since the $(\nu_n : n \in \mathbb{N})$ sequence satisfies the LDP with rate function $I$, Varadhan's theorem (see Ellis \cite[Th. II.7.1.b]{1985ELDSM_Ellis}) states that if
\begin{equation}
  \label{eqn:suffConditionVaradhan}
  \lim_{M \rightarrow \infty} \limsup_{n \rightarrow \infty} \frac{1}{n} \ln \int_{t \geq \frac{M}{\beta}} \exp\{ n \beta t \} ~d\nu_n(t) = -\infty
\end{equation}
then the limit
\begin{equation}
  \label{eqn:varadhan}
  \lim_{n \rightarrow \infty} \frac{1}{n} \ln \int_{\mathbb{R}} \exp\{n\beta t\} ~\nu_n(dt) = \sup_{t \in \mathbb{R}} \left\{ \beta t - I(t) \right\}
\end{equation}
holds. The integral on the left side in (\ref{eqn:varadhan}) can be simplified by defining the finite cardinality set
\[
  A_n=\{- n^{-1} \ln P_n(x^n): \forall x^n \in \mathbb{X}^n\} \subset \mathbb{R}
\]
and by observing that
\begin{eqnarray*}
  \lefteqn {\int_{\mathbb{R}}\exp\{n\beta t\}~\nu_n(dt) } \\
&& =\sum_{t \in A_n}\exp\{n\beta t\}\sum_{x^n: P_n(x^n)=\exp\{-nt\}}P_n(x^n) \\
&& =\sum_{x^n}P_n(x^n)^{1-\beta} \\
&& =\sum_{x^n}P_n(x^n)^{\frac{1}{1+\rho}} = \exp \left\{ \beta H_{1/(1+\rho)}(P_n) \right\}.
\end{eqnarray*}
Take logarithms, normalize by $n$, take limits, and apply (\ref{eqn:varadhan}) to get the desired result. It therefore remains to prove (\ref{eqn:suffConditionVaradhan}).

The event $\{ t\geq \frac{M}{\beta} \}$ occurs if and only if $$\left\{ P_n(x^n) \leq \exp\left\{\frac{-nM}{\beta} \right\}\right\}.$$ The integral in (\ref{eqn:suffConditionVaradhan}) can therefore be written as
\begin{eqnarray*}
  \lefteqn{ \sum_{t \in A_n, t\geq \frac{M}{\beta}} ~~ \sum_{x^n: P_n(x^n)=\exp\{-nt\}}\exp\{n\beta t\}P_n(x^n) } \\
    & = & \sum_{x^n:P_n(x^n)\leq \exp \{\frac{-nM}{\beta}\}} P_n(x^n)^{\frac{1}{1+\rho}}\\
    & \leq & |\mathbb{X}|^n \cdot \exp\Big\{\frac{- n M}{\beta(1+\rho)}\Big \}.
\end{eqnarray*}
The sequence in $n$ on the left side of (\ref{eqn:suffConditionVaradhan}) is then
\[
  \ln |\mathbb{X}|-\frac{M}{\beta(1+\rho)},
\]
a constant sequence. Take the limit as $M \rightarrow \infty$ to verify (\ref{eqn:suffConditionVaradhan}). This concludes the proof.


\section{Conclusion}
We first showed that the problem of finding the limiting guessing exponent is equal to that of finding the limiting compression exponent under exponential costs (Campbell's coding problem). We then saw that the latter limit exists if the sequence of distributions of the information spectrum satisfies the LDP (sufficient condition). The limiting exponent was the Legendre-Fenchel dual of the rate function, scaled by an appropriate constant. It turned out to be the limit of the normalized cumulant of the information spectrum random variable. While some of these facts can be gleaned from the works of Pfister \& Sullivan \cite{200411TIT_PfiSul} and Merhav \& Arikan \cite{Arikan-Merhav}, our work sheds light on the key role played by the information spectrum. It will be of interest to find a rich class of sources beyond those listed in this paper for which the information spectrum satisfies the LDP.

Results on guessing with key-rate constraints for a general source are provided using the above information spectrum approach in \cite{200901TRPME_HanSun}.

\bibliographystyle{IEEEtran}
{
\bibliography{IEEEabrv,wisl}
}

\end{document}